\documentclass[twocolumn]{aastex631}

\shorttitle{Superflares from Recurrent Nova V2487 Oph}
\shortauthors{Schaefer}
\graphicspath{{./}{figures/}}

\begin{document}
\title{Recurrent Nova V2487 Oph Had Superflares in 1941 and 1942 With Radiant Energies 10$^{42.5\pm1.6}$ Ergs}

\author[0000-0002-2659-8763]{Bradley E. Schaefer}
\affiliation{Department of Physics and Astronomy,
Louisiana State University,
Baton Rouge, LA 70803, USA}



\begin{abstract}

V2487 Ophiuchi (V2487 Oph) is a recurrent nova with classical nova eruptions in 1900 and 1998, and it is also the most extreme known superflare star.  These superflares are roughly-hour-long flares with amplitudes and optical energies reaching up to 1.10 mag and $10^{39.21}$ ergs, with the superflares recurring once-a-day.  The V2487 Oph superflares are certainly operating with the same mechanism as all the other types of superflare stars, where magnetic loops are twisted and stretched until reconnection occurs, whereupon ambient electrons are accelerated to relativistic energies and then emitted bremsstrahlung radiation from X-ray to radio.  V2487 Oph is unique among known superflare stars in that one of the loop footprints is in an accretion disk.  This exact mechanism was theoretically predicted by M. R. Hayashi and colleagues in 1996.  Now, I have found two superflares recorded on Harvard archival photographs from the years 1941 and 1942.  These two superflares have $B$ magnitude amplitudes of $>$1.83 and $>$2.00 mag and total radiated energies of $10^{42.4}$ and $10^{42.5}$ ergs with bolometric corrections.  Each has emitted energies of $\sim$30-billion Carringtons, in units of the most energetic solar flare.  Further, I find superflares in the {\it Zwicky Transient Factory} light curves, so V2487 Oph has been superflaring from 1941 to 2023.  For the observed number distribution of $dN/dE$=$4E^{-2}$ superflares per year, for $E$ in units of $10^{41}$ ergs, the emitted energy in superflare light is $10^{42.1}$ erg in each year, or $10^{44.1}$ ergs from 1941 to 2023.

\end{abstract}


\section{Introduction}

Superflare stars were discovered as a new and separate phenomenon back in 1989, when a collection of otherwise-ordinary stars were recognized as having short duration flares that are of enormous energies (Schaefer 1989).  I defined the superflares to have energies $>$1000$\times$ those of even the largest solar flares and stellar flares from flare stars, so as to constitute a separate class of phenomena (Harwit 1984).  With the largest energy for solar flares\footnote{The Carrington event of 1859 has estimates ranging from 2$\times$10$^{30}$ to 4$\times$10$^{32}$ erg (Tsurutani et al. 2003; Hudson 2021).  The historical `Miyake Events' from $^{14}$C anomalies in tree rings dated to 774 and 994 AD (Miyake et al. 2012; 2013) have large uncertainties in energy, but are ``within the same order of magnitude as that of the Carrington event'' (Hudson 2021).  The energy of $10^{32}$ ergs can be taken as a new unit named the `Carrington', with this being convenient and evocative.} being around 10$^{32}$ ergs, this makes superflares to have energies $>$10$^{35}$ ergs.  Superflares are seen on stars that are indistinguishable from our Sun, including several of the closest known solar twins (Schaefer 1989; Schaefer, King, \& Deliyannis 2000).  This new class of variables has been confirmed with many excellent light curves from the {\it Kepler} and {\it TESS} spacecrafts (Maehara et al. 2012; 2015; Shibayama et al. 2013; Nogami et al. 2014; Tu et al. 2020)  Attention has been concentrated on these Sun-like superflare stars, with implications for the retention of exoplanet atmospheres, with implications for the origin of life, with implications for the possibility of superflares on our own Sun, plus implications for the habitability of planets (e.g., Lingam \& Loeb 2017; Yamashiki et al. 2019).

Little attention is given to the fact that superflares are seen all across the HR diagram.  I have collected lists of superflares on B and A main sequence stars, on giant stars, on supergiants, on white dwarfs, on very-low mass main sequence stars, and on Mira stars, (Schaefer 1989; 1990; 1991).  These classes of superflares have been confirmed by many later studies (de Laverny et al. 1998; Mullan 2009; Balona 2011; 2012; Craine et al. 2015; Howard et al. 2018; Kielkopf et al. 2019).  Further studies have extended the superflare phenomena to brown dwarfs, extreme horizontal branch stars, and variable stars of most types (Gizis et al. 2017; Momany et al. 2020; Zhong et al. 2023).  Importantly, large numbers of superflares have been measured with wonderful light curves with the {\it Kepler} and {\it TESS} spacecrafts, and these turned the whole field from a data-poor study into a data-rich field.  These studies further demonstrate that superflares occur frequently among stars all across the HR diagram (Balona 2011; 2012; Althukair \& Tsiklauri 2023)  One implication is that the superflare mechanism is universal, applicable to stars of all types.

The superflare mechanism is the same mechanism as for solar flares and stellar flares on ordinary flare stars.  That is, magnetic loops with footprints on the star's surface are twisted and stretched until they suffer the poorly-understood process of magnetic reconnection.  With the reconnection, ambient electrons are accelerated by Alfv\'{e}n waves to relativistic velocities along the magnetic loops, where they impact onto the star surface such that bremsstrahlung radiation sends out a continuum of photons from X-ray to radio.

Superflare properties are uniform across all classes.  Their light curves are all fast-rise with exponential decay, often with short impulsive spikes near the beginning.  Typical durations are hours, with the durations scaling as a power of the flare energy, with $E^{0.4}$.  The low-energy flares are common while the high-energy flares are rare, with the number distribution ($dN/dE$) scaling as $E^{-2.0\pm0.2}$.  The time from one superflare event to the next is proportional to the energy of the first event ($\Delta T$$\propto$$E$).  Despite the many nice superflare light curves, I am aware of no optical spectra of superflares, at least up until last year for V2487 Oph.

\section{RECURRENT NOVA V2487 OPH}

Nova Oph 1998 was discovered on 1998 June 15 by K. Takamizawa (Saku-machi, Nagano-ken, Japan) at magnitude 9.5, and was designated as V2487 Oph.  The full light curve is collected in Schaefer (2010), with a peak at $B_{\rm peak}$=10.1 and $V_{\rm peak}$=9.5, plus a fast decline by 3 mag in 8.4 days that flattened out to a plateau, before quickly going to quiescence of $B_{q}$=18.1 and $V_{q}$=17.3.  Its outburst spectrum showed high-energy He {\rm II} lines and a FWHM velocity of 10,000 km s$^{-1}$ (Lynch et al. 2000).  The very fast $t_3$=8.4 days, the small amplitude of 7.8 mag in the V-band, the P-class light curve shape, the high-excitation lines, and the extremely high expansion velocity are all pointing strongly to the system being a recurrent nova (RN, Hachisu et al. 2002; Pagnotta \& Schaefer 2014).

A. Pagnotta (now at the College of Charleston) made a full search of the Harvard Observatory collection of archival astronomical sky photographs.  These photographs are from emulsion coated onto one side of a thin glass plate, usually 8$\times$11 inches in size, with stars appearing as black spots on a clear sky background.  The Harvard archives are a collection of nearly half-a-million plates covering the entire sky from 1889--1989.  This coverage\footnote{The only significant gap in the 101 year interval is the Menzel Gap from 1954 to 1969 caused by the then-Director of the Observatory, D. Menzel, defunding the entire plate program so as to fund a solar eclipse expedition of his own.} is typically thousands of plates for all stars brighter than $B$=12, and hundreds of plates for all stars brighter than $B$=15.  With this coverage, there was reasonable hope of catching a long-ago RN event from V2487 Oph.

Pagnotta et al. (2009) discovered a long-lost eruption from the year 1900.  The eruption was imaged on only one plate, numbered as AM 505, a 61 minute exposure on 1900 June 21 with a limiting magnitude of $B_{\rm lim}$=11.3.  The nova appears as a star image with magnitude $B$=10.27$\pm$0.11.  The new star image is highly significant, and at the exact location of V2487 Oph.  Importantly, the point-spread-function (PSF) of the image was exactly like the PSFs of nearby stars, in this case being a star trail 1.8 arc-minutes long brighter at both ends to form a short dumbbell shape.  (Likely, this is caused by the ordinary periodic error of the drive.)  This characteristic PSF shape is identical with those of nearby stars, hence providing the proof that the light recorded on the plate came from a real stellar source far outside the Solar System.  So V2487 Oph is an RN, with eruptions at least in 1900 and 1998.

That is a recurrence timescale of 98 year.  But the very short duration and the extreme expansion velocity imply that the real recurrence timescale is greatly shorter than 98 years, hence there are likely several missed eruption between 1900 and 1998.  Pagnotta et al. (2009) estimate that the real recurrence timescale is 18 years, with substantial uncertainty, so that roughly four or five eruptions have been lost to observers over the past century.

The most important parameter for any cataclysmic variable (CV) is the orbital period of the underlying binary.  V2487 Oph does not eclipse, nor show any obvious photometric modulation.  With a massive effort using ground-based time series, a weak and transient photometric periodicity of 1.24$\pm$0.02 days was suggested (Schaefer, Pagnotta, \& Zoppelt 2022).  In addition, the $K2$ mission of the {\it Kepler} spacecraft produced a 69-day light curve that displayed a periodicity of 1.2216$\pm$0.0031 days.  This signal only appeared in the first 25 days and only in the superflare light.  For the remaining time, no periodic signal is seen in the non-flare light, while the flare light from 36 superflares beat together to give weak signals, for example at 1.6 days (Rodr\'{i}guez-Gil et al. 2023).  I conclude that the 1.24 day periodicity is not orbital, and that the flare light does not have any sort of a coherent periodicity, rather the 1.24 days represents an average recurrence timescale.  Convincingly, the orbital period was found from a nice radial velocity curve to be 0.753$\pm$0.016 days (Rodr\'{i}guez-Gil et al. 2023), and this solves the issue of the orbital period.  With the new orbital period, the companion star must have a radius near 0.96 R$_{\odot}$, which makes it into an early G-type main sequence star (if not evolved), just like our Sun.  While the orbital period is reliable, the measured $\gamma$ and $K$ velocities vary greatly from line-to-line, while the tentative mass of the companion star (0.21 M$_{\odot}$) is not reasonable, so unrecognized complications (for example, if the emitting material possesses a significant vertical velocity component) must be distorting the observed radial velocity curves.

The second most important parameter for any CV is the distance.  The {\it Gaia} parallax is 0.129$\pm$0.075 milliarcseconds, which is consistent with a distance like that to the galactic center.  With an angular distance of 10.2$\degr$ from the Galactic center, V2487 Oph is confidently part of the bulge population, which for novae has a distribution with a central distance of 8000 pc and a Gaussian radius of 750 pc.  With a full Bayesian calculation including all available information, the best distance to V2487 Oph is 7540$\pm$730 pc (Schaefer 2022).  With this and the measured $E(B-V)$=0.5$\pm$0.2, the peak absolute magnitude, $M_{V_{\rm peak}}$, becomes -6.44$\pm$0.66, (Schaefer 2022), which is similar to most other novae and RNe.  In quiescence, the absolute visual magnitude is $M_{V_{q}}$=1.4, almost entirely from the accretion disk (Schaefer et al. 2022).

In quiescence, V2487 Oph is an extreme-ultraviolet source as seen by {\it ROSAT} (Voges et al. 1999), an X-ray source as seen by {\it XMM-Newton} (Hernanz \& Sala 2002), and a $\gamma$-ray source as seen by {\it INTEGRAL} (Krivonos et al. 2010).  This all emphasizes the unique nature of V2487 Oph, greatly different from almost all other novae or CVs.  Unfortunately, there is not enough information to attribute the high-energy flux to either the quiescent system or to superflares.

The nature of the accretion flow will become important for models of superflares.  The accretion rate must be very high (up near the steady hydrogen burning limit) so as to produce the fast recurrence timescale.  Detailed measures and arguments return a range of (3--16)$\times$10$^{-8}$ M$_{\odot}$ yr$^{-1}$ (Schaefer et al. 2022).  The accretion must be by Roche lobe overflow as the only way to be so large.  The SED shows a full disk component with no high-frequency cutoffs, so this points to V2487 Oph not having its disk significantly truncated by the white dwarf magnetic field\footnote{The spectral energy distribution shown in figure 5 of Schaefer et al. (2022) shows no cutoff of the disk flux at the highest frequencies, in particular, the $GALEX$ brightness is within a few per cent of the fit for an $\alpha$ disk.  In the $GALEX$ band, one-quarter of the flux comes from the inner 3$\times$10$^9$ cm of the disk.  So the inner disk cannot be truncated to any radius larger than 0.04 $R_{\odot}$.  This can be compared to the white dwarf radius of 0.005 $R_{\odot}$ and to the outer radius of the disk, typified as the circularization radius, of 0.5 $R_{\odot}$.  The point is that the white dwarf magnetic field is dynamically negligible for almost all of the disk, including where the model loop footprints are anchored.}.  However, the X-ray spectra are similar to those of Intermediate Polars, so perhaps the disk is truncated within its inner 8\% from magnetic effects (Rodr\'{i}guez-Gil et al. 2023).

For the last 18 years, J. Grindlay (Harvard University) has been leading a massive effort to completely digitize all of the direct image plates at Harvard.  His vision was that the Harvard archives must be scanned to completion, as this enables comprehensive large-scale studies of whole classes of stars, as well as pointed studies for any specific target of interest.  The Harvard plates are unique in covering a full century of history, with this precious resource being the only way to get the long-term record of most stars in the sky.  Such extensive and exhaustive programs are effectively impossible by the traditional by-eye searches.  Further, a fully scanned digital archive would become available to the entire world, not just the few living experts who travel to Cambridge, thus greatly expanding and democratizing their availability.  With the current nearly-complete loss of expertise and knowledge in extracting science from the plates, Grindlay has also lead the effort to use the scans to calculate accurate photometry for every star image for every plate.  The resultant software (Tang et al. 2013) produces excellent photometry, equally as good as my own by-eye measures in most cases.   Grindlay's program is named {\it Digital Access to a Sky Century @ Harvard}, or simply DASCH\footnote{\url{https://dasch.cfa.harvard.edu/}}.  With this, DASCH is the only plate archive in the world that has their plates being scientifically useful to workers other than experts.  One month ago, DASCH completed the vast task of scanning all of Harvard's direct image plates.

For the case of V2487 Oph, with the effective completion of DASCH, I realized that there is an opportunity that goes past the old by-eye search for lost RN eruptions.  In particular, DASCH has digitized 3406 plates that show the V2487 Oph field, and this is greatly more than was practically possible for the by-eye searches.  Most of the added magnitudes are from AI, BI, and FA plates of low quality, or from plates with the target near the edge with a poor PSF.  These plates added to the prior by-eye searches have limiting magnitudes typically of $B$=11 mag, so they could only catch V2487 Oph in eruption for one or two days at peak.  Still, the large number of such plates could fill in gaps in the light curve that might happen to include a long-lost eruption.  Further, DASCH shows that 14 very deep plates from the BR series are available from the years 1941 and 1942.  This series was missed by the old by-eye searches because it covers only small areas of the sky and for only two years.  Out of all 3406 plates, DASCH reports positive detections for 42 plates.  On my by-eye examination, all of these detections are just ordinary plate artifacts that confuse the DASCH software, or are ordinary insignificant random grain enhancements.  The two exceptions to this are the plates AM 505 (with the 1900 eruption) and BR 2506 (see below).  In the end, with the DASCH search for long-lost eruptions, none were found.

\section{SUPERFLARES ON V2487 OPH}

In 2015, A. Pagnotta and I proposed for short cadence observations with Cycle 9 of the $K2$ mission (on the {\it Kepler} spacecraft), seeking photometric modulations on the orbital period.  Our accepted proposal (GO 9912, PI Pagnotta) resulted in 67 days of a nearly continuous light curve with 59 second time resolution, all starting in April of 2016.  Instead of periodic modulations, we discovered 60 highly-significant flares, each isolated between intervals of a constant brightness in quiescence (Schaefer et al. 2022).  All the light curves consist of a fast-rise (typically under 6 minutes) and exponential-decline, always with one-to-three spikes near the start.  The flare light curves have a median duration of just under one hour, with the fading tails extending longer.  The flare amplitudes were observed up to 1.10 mag, and the calculated optical energies, $E$, are as high as $10^{39.2}$ ergs.  Further, 11 flares were found in the archives of the Zwicky Transient Factory (ZTF).

The V2487 Oph flares obey four rules (Schaefer et al. 2022):  First, the number distribution of flares varies as $dN/dE$$\propto$$E^{-2.34\pm0.35}$.  Second, the waiting time from one flare to the next scales as $\Delta T$$\propto$$E$.  Third, the flare durations scale as $D$$\propto$$E^{0.44\pm0.03}$.  Fourth, the flare light curves consist of a fast-rise exponential-decline shape with initial spikes.  This suite of four properties is exactly the same as had already been previously described as the properties of all the classes of superflare stars (Schaefer et al. 2022, table 7).  This suite of four properties is denied by all models and precedents for scenarios where the flare energy comes from nuclear burning on the white dwarf or from the release of gravitational energy of material falling onto the white dwarf (Schaefer et al. 2022, table 7).  This is a proof that V2487 Oph is suffering from superflares  (Schaefer et al. 2022).  By extension, this means that these superflares are from the mechanism of magnetic loops getting stretched and tangled, leading to magnetic reconnection, resulting in relativistic electrons raining down on the loop footprints, with these producing a flash of bremsstrahlung radiation from the radio to the X-ray.

Rodr\'{i}guez-Gil et al. (2023) obtained spectra of one superflare on V2487 Oph.  These spectra from 3000--25000~\AA~were taken on the night of 2019 June 8 with the X-shooter echelle spectrograph at the 8.2-m VLT in Cerro Paranal, Chile.  The flux from the flare is almost all continuum light, with this bright in the near-ultraviolet and near-zero in the near-infrared.  On top of this very blue continuum, the added flare light has emission lines from H$\alpha$ plus other Balmer lines, He {\rm II} $\lambda$4686, the Bowen blend, and the Ca {\rm II} triplet.  This flare light extends from velocities of $-$2000 to $+$1200 km s$^{-1}$, with some variability perhaps as a function of orbital phase, so the superflare is ejecting substantial amounts of mass at high velocity.

V2487 Oph is the only known superflare star in a close interacting binary.  V2487 Oph was already the most extreme superflare star, with by far the fastest recurrence time of nearly once-per-day, and with the highest known radiant energy.  It cannot be a coincidence that the most extreme superflare star is also the only known case in an interacting binary.  So there must be some causal connection for why the magnetic reconnection in this accreting binary results in the most-energetic and most-frequent superflares.

Schaefer et al. (2022) proposed the scenario where the magnetic loops connect the companion star to material in the accretion disk.  (For ordinary superflares on single stars, both footprints are on the star.)  This scenario requires loop-lengths that are greatly longer than possible for a single star, and the superflare energy is proportional to the loop-lengths, hence a partial explanation of why the V2487 Oph events must be at the top end of flare energy.  This scenario requires that the footprint in the rotating accretion disk provide rapid stretching, with the stretching of the magnetic field making for amplification of the field strength, hence the rest of the explanation for why the high energy can be quickly supplied to the field lines.  This scenario requires that the loops in the accretion disk be twisted by the differential rotation in the loop footprints injecting helicity and snarled with other loops, hence an explanation for how major reconnections can be triggered as fast as once-per-day.  Once the field lines are connected from the companion into the accretion disk, the field strength will be speedily increased, twisted to reconnection, speedily increased again, twisted to reconnection for another superflare, and so on.  So we have a simple explanation for why V2487 Oph is the most extreme superflare source.

Unknown to me until a few months ago, Hayashi, Shibata, \& Matsumoto (1996) had previously published a theory paper that presents a model that is closely the same as the schematic picture presented in the previous paragraph.  Hayashi et al. give a detailed physics model and calculate quantitative predictions.  In essence, they made a {\it prediction} of exactly what we see for superflares on V2487 Oph.  Their specific model is for protostars, but this is irrelevant because all they are using is a star and a disk, with closed magnetic loops connecting the two.  ``The closed magnetic loops connecting the central star and the disk are twisted by the rotation of the disk.  As the twist accumulates, magnetic loops expand and finally approach the open field configuration.  A current sheet is formed inside the expanding loops. In the presence of resistivity, magnetic reconnection takes place in the current sheet.  The timescale of this `flare' is the order of the rotation period of the disk.'' (Hayashi et al. 1996).  For V2487 Oph, the rotation period of the disk  at its outer edge (typified by the circularization radius) is just under one hour, thus matching the duration of the superflares.  They demonstrate that the magnetic fields keep increasing linearly with time, until reconnections occur.  They also demonstrate the physics that the injection of helicity by the rotation of the footprint in the disk will twist the field lines and necessarily lead to reconnection.  The reconnection event will eject hot plasma with velocities of 200--400 km s$^{-1}$ for the particular case they looked at.  The flare emission extends up to hard X-rays.  In the end, Hayashi et al. predicted superflares for the case of a star next to an accretion disk, with properties that are the same as for V2487 Oph.  Thus, V2487 Oph provides a  good confirmation of insightful theory predictions from 1996.  And the Hiyashi et al. paper provides a confirmation that the V2487 Oph superflares are indeed superflares with magnetic reconnections for loops connecting the star and the disk.

\section{SUPERFLARES IN 1941 AND 1942}

After being alerted by DASCH to the coverage afforded by the 14 excellent BR plates from 1941--1942, I examined all of them by-eye on the original glass photographs.  This by-eye examination is the century-long traditional method, as used by the all the great researchers at Harvard, including H. Leavitt, A. J. Cannon, and C. Payne-Gaposchkin.  In practice, this entails placing the glass onto a light table and examining the star images with an 8$\times$ loupe.  This allows the fine details of the PSF to be seen, and the magnitudes are measured effectively by looking at the image's radius out to some middling isophotal level.  When looking at the images, the eye can easily pick out a constant isophotal level from star-to-star, and compare the radius of the target star versus the radii of comparison stars of known magnitude.  A simple interpolation of the isophotal radius of the target star with the radii of the comparison star returns an accurate magnitude.  The spectral sensitivity of the Harvard plates is close to that defining the modern B-magnitude system, and the comparison star magnitudes are calibrated from the B-magnitudes from the APASS program\footnote{AAVSO Photometric All-Sky Survey (APASS) is the source for getting good classic $B$ magnitudes for most stars in the sky down to fainter than 17th mag or so.  See \url{https://www.aavso.org/download-apass-data}.  It is important that the comparison stars be in the same system for both the Harvard and modern data, while the APASS is the only general source that is closely in the modern $B$ system.}.  Thus, the resultant magnitudes are accurately in the modern Johnson B-magnitude system.  In practice for V2487 Oph, the total photometric uncertainty is typically 0.15 mag.  For the circumstances of V2487 Oph, the by-eye magnitudes are substantially better than are those produced by DASCH.

For the 14 BR plates, two have highly significant detections of the RN at brightnesses far brighter than the mean quiescent level of $B_q$=18.1.  Plate BR 2506 is a 45 minute exposure taken on the night of 1941 July 25--26, and it shows the RN at a $B$ magnitude of 16.81$\pm$0.11 (see Fig. 1).  Plate BR 3244 is a 45 minute exposure taken on the night of 1942 July 10--11, and it shows V2487 Oph at $B$=16.64$\pm$0.12 (see Fig. 2).  In both cases, the RN image is at exactly the correct position on the plate, the image is highly significant, and the image passes my various tests to detect any of the variety of artifacts.  Importantly, the target PSFs are identical to the PSFs of nearby stars of similar brightness.  This provides the proof that these two images are confident detections of V2487 Oph.

\begin{figure}
\epsscale{1.15}
\plotone{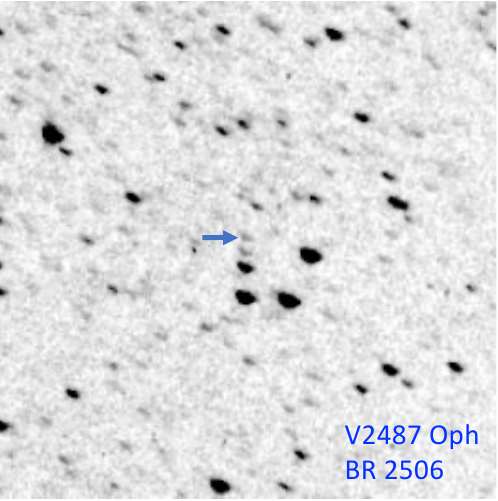}
\caption{The 1941 superflare.  The position of V2487 Oph is at the position of the highly-significant star image indicated by the blue arrow.  Critically, the shape of the point-spread-function for the target star is the same as those of nearby stars of similar brightness.  This provides the proof that the target image is a recording of a point source on the sky, and not some artifact or passing asteroid.  The identification of several of the APASS comparison stars are given in Fig. 3.  With this, we can see that the 1941 superflare is somewhat fainter than the B=16.62 star and substantially brighter than the B=17.05 star.  My measured brightness for V2487 Oph is 16.81$\pm$0.11 on the modern $B$ magnitude system.  This image is a 5$\times$5 arc-minute square extracted from the DASCH scan of plate BR 2506, with north to the top, and east to the left.  This plate is a 45 minute exposure on the night of 1941 July 25--26, made with the 8-inch refractor at Bloemfontein in South Africa. }
\end{figure}

\begin{figure}
\epsscale{1.15}
\plotone{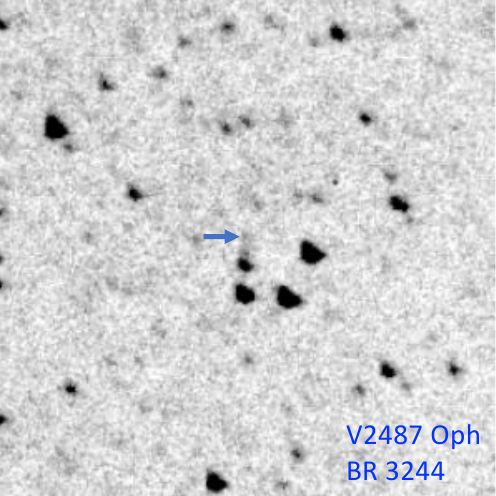}
\caption{The 1942 superflare.  The blue arrow points at the position of V2487 Oph, where there is a highly-significant star image.  Critically, the PSF for the target star is the same as those of nearby stars of similar brightness, so the target image was formed by a star, and not some artifact or passing asteroid.  In  comparisons with stars of known $B$ magnitude (see Fig. 3), the 1942 superflare is fairly close to the B=16.62 star.  My measured brightness is $B$=16.64$\pm$0.12.  This image is a thumbnail from the DASCH scan of plate BR 3244, with 5 arc-minutes on the sides, north to the top, and east to the left.  This plate is a 45 minute exposure on the night of 1942 July 10--11. }
\end{figure}

These two detections are certainly not the RN in any sort of nova eruption.  The proof comes because both plates have deep plates immediately-before and immediately-after that do not show the star to limits substantially fainter than the positive detections.  For BR 2506, we have a plate taken 3 nights earlier that shows nothing to a limit of 17.51 mag, and we have two plates taken 3 nights later that shows nothing to 16.92 mag and to 17.6 mag.  For BR 3244, a plate taken on the previous night shows the RN to be $B$$>$18.0 (see Fig. 3).

The observed brightness levels for the two plates ($B$ measures equal to 16.81 and 16.64 ) are far above the normal flickering as seen in the many modern time series.  The normal quiescent level is olden times is shown to be fainter than $B_{\rm lim}$=18.0 by the various deepest Harvard plates.  An example is displayed in Fig. 3, where the nova position is empty to a limiting magnitude of 18.0.  In this figure, I have labeled four comparison stars with left-pointing arrows and their $B$ magnitudes from APASS.  The point of this is to demonstrate that V2487 Oph had a typical quiescent magnitude in the 1940s that is consistent with the modern average of 18.1, and thus that the two BR plates {\it do not} record the RN in some flicker or high state.

\begin{figure}
\epsscale{1.15}
\plotone{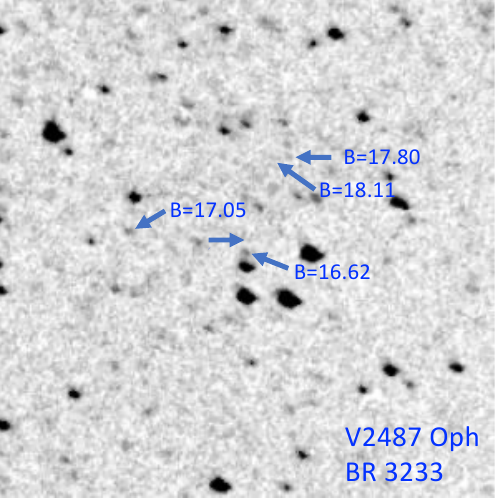}
\caption{V2487 Oph in quiescence.  The blue right-pointing arrow points at the position of V2487 Oph, where the nova is conspicuously missing.  This plate (BR 3233) is from 1942 July 9--10, just one night before the superflare shown in Fig. 2.  Four APASS comparison stars are indicated with left-pointing arrows and their $B$ magnitudes.  The $B$=18.11 star appears faintly as an insignificant cluster of grains.  V2487 Oph is certainly fainter than 17.80, and I place the limit as $B$$>$18.0.  This demonstrates that the RN was not in any sort of a high state at the time of the superflares, but rather at a level consistent with the $B_q$=18.1 from the last two decades.  Like the other figures, this figure is 5 arc-minutes on a side, with north at the top and east to the left. }
\end{figure}

The two bright BR plates in 1941 and 1942 cannot arise from a nova event of any type, nor from any sort of flickering or state-changes.  And the duration of these observed events is under one day, so they can be correctly termed as `flares'.  These flares have the strong precedent of the superflares on V2487 Oph, as observed with high frequency by $K2$ and the ZTF.  So I conclude that the Harvard plates have recorded two superflares back in 1941 and 1942.

What are the amplitudes of these superflares?  The simplistic calculation is 18.1-16.81=1.29 and 18.1-16.64=1.46 mag for the two plates.  But this is certainly bad, because amplitude involves the peak magnitude, rather than an average over a random 45 minutes time interval.  The $K2$ superflares show a peaked impulsive phase typically lasting 10--20 minutes.  This impulsive phase will make the true maximum light to be substantially brighter than the 45 minute average recorded on the plate.  So the real amplitudes of the two superflares are certainly much greater than 1.29 and 1.46 mags.  The real peaks will depend on the exact shape of the light curve, which varies substantially from flare-to-flare.  For the eight largest-amplitude $K2$ superflares, I have calculated the difference in amplitude, in magnitudes, between the peak brightness and the optimal 45 minute average.  The average difference is 0.68 mag, with a range of 0.54--0.80 mag.  So we can safely say that the amplitude is $>$0.54 mags greater than in the above simplistic calculation.  This correction in amplitude should actually be actually larger, because the Harvard 45 minute exposures are unlikely to be optimally timed to record the maximum flux.  With this, we know that the real amplitude of the 1941 superflare was $>$1.83 mag, while the real amplitude of the 1942 superflare was $>$2.00 mag, and likely substantially larger.

The largest amplitude for any of the 60 $K2$ superflares was 1.10 mag, while the well-observed ZTF flare had an amplitude of 0.36 mag, while the single ZTF 30 second exposures all have magnitudes within 0.98 mag of its quiescent level.  With the two Harvard superflares having amplitudes $>$1.83 and $>$2.00 mag, these are now by far the most energetic superflares of any known for V2487 Oph, as well as the most energetic superflare known for any star.

\section{SUPERFLARE ENERGIES}

How much radiant energy is emitted during these two superflares?  Our only specific input for these flares is the observed magnitude as averaged over the 45 minute integrations.   

First, we should calculate the absolute magnitude for the flare light alone.  The observed $\langle B \rangle$ values are 16.81$\pm$0.11 and 16.64$\pm$0.12 mag for the 1941 and 1942 superflares respectively.  For the best distance of 7540$\pm$730 pc and the $B$-band extinction of  $A_B$=4.1$\times$$E(B-V)$=2.05$\pm$0.82 value, the offset from $\langle B \rangle$ to absolute magnitude is -16.44$\pm$0.85.  The $\langle M_B \rangle$ for the flare+quiescent light is $+$0.37$\pm$0.86 for the 1941 superflare and is $+$0.20$\pm$0.86 for the 1942 superflare.  The absolute $B$-magnitude of the system in quiescence, with $B_q$=18.1, is $M_{B_q}$ is $+$1.66.  This quiescent flux can be subtracted from the flare+quiescent flux by means of the usual magnitude equation relating to flux.  With this, $M_{B}$ is $+$0.77$\pm$0.86 for the 1941 superflare and $+$0.54$\pm$0.86 for the 1942 superflare, as averaged over the 45 minute integrations.

A question might arise as to whether any systematic effects are caused by the non-linearities of the photographic process.  In particular, most of the star images are saturated in their centers.  Further, the relation between the photographic density and the incoming light flux (the H-D curve) is approximated as a power law only over a narrow range, so the unsaturated emulsion response is highly non-linear.  With strong experience from CCD photometry, the saturation and non-linearity would appear to preclude any good photographic photometry.  Nevertheless, these properties of all emulsions are irrelevant for real photographic photometry.  The reason is that the magnitude estimates are essentially measuring all star images out to the same isophotal level, so the critical position on the H-D curve is just some constant all around each star image and the same constant from the target to the comparison stars.  The emulsion's non-linearity has zero effect because all measures are all out to the same constant isophot.  As the flux from a star increases, the fiducial isophot is pushed farther out in the tail of the PSF.  The measured isophots are for the ring in the emulsion where the star image is far from saturation.  So the answer to the question of saturation and non-linearities is that these have zero effect on the photographic photometry.

The second calculation needed to derive the superflare energies is to convert the $M_B$ values into bolometric absolute magnitudes, $M_{\rm bolo}$.  The $M_B$ value is for the luminosity of the superflare light as integrated over the $B$ passband, while the $M_{\rm bolo}$ value is for the luminosity of the superflare light as integrated over all photon frequencies.  The bolometric correction is $M_{\rm bolo}$-$M_B$, here defined as a quantity that is always negative.  This correction depends on how much of the superflare energy is radiated blueward and redward of the $B$ passband.  For the superflares of V2487 Oph specifically, Rodr\'{i}guez-Gil et al. (2023) measured the spectral energy distribution (SED) of the superflare light from the ultraviolet to the middle-infrared (0.3--2.5 $\mu$m), and it shows a very blue continuum.  From their data, the continuum spectral flux scales as $f_{\nu} \propto \nu ^{0.10\pm0.15}$.  V2487 Oph is an extreme-UV source (Voges et al. 1999), an X-ray source (Hernanz \& Sala 2002), and a $\gamma$-ray source (Krivonos et al. 2010), but it is unclear whether the high energy flux is from the superflares or the quiescent binary.  Nevertheless, all classes of superflares are known to have the flaring flux extending brightly out to the X-rays, so it is likely that V2487 Oph has its superflare flux extending out to the X-ray regime.  The superflare light is coming from the brehmsstrahlung radiated by the large swarms of electrons accelerated to relativistic energies that impact into the accretion disk and star surface.  For frequencies below the exponential cut-off, the bremsstrahlung SED scales as $f_{\nu} \propto \nu ^{0}$.  Given the precedents of all other superflares being strong X-ray sources and the relativistic electrons that must be in the very large magnetic loops of V2487 Oph, the cut-off needs to be in the X-ray regime, likely from 1--1000 keV.  So, for calculating the bolometric correction, we have a clear case for an SED scaling as $\nu ^{0.10\pm0.15}$ and extending out to at least the X-ray regime.  For a cut-off at 1 keV, the bolometric correction is -8.5 mag.  The primary uncertainty is the frequency of the cut-off.  We can take the most conservative case to be for an effective cut-off at 0.1 keV, and for the SED scaling as $\nu ^{-0.05}$, for a bolometric correction of $-$5.3 mag.  With this limit on the correction, $M_{\rm bolo}$ averaged over the 45 minute integrations is $<$-4.5 and $<$-4.8 for the flares in 1941 and 1942.

The third calculation is to convert the bolometric absolute magnitudes into luminosities, with units of ergs per second.  The usual way is to scale from our Sun's known values of $L_{\odot}$=3.83$\times$10$^{33}$ erg s$^{-1}$ and $M_{{\rm bolo},{\odot}}$=4.75.  The scaling is 
\begin{equation}
L = L_{\odot} 10^{-0.4(M_{{\rm bolo}}-M_{{\rm bolo},{\odot}})}.
\end{equation}
With this, the luminosity is $>$1.9$\times$10$^{37}$ erg s$^{-1}$ for the 1941 superflare and $>$2.5$\times$10$^{37}$ erg s$^{-1}$ for the 1942 superflare, still as averaged over the 45 minutes of the individual Harvard exposures.  This is getting near the Eddington limit for a white dwarf (and the peak superflare luminosity gets even closer), but that is fine as we have material at velocities up to 2000 km s$^{-1}$ (Rodr\'{i}guez-Gil et al. 2023) in what appears to be an explosion that has accelerated gas to far above the escape velocity.

The fourth calculation is to get the radiant energy from the luminosity averaged over the 45 minute exposures.  This correction is just a factor 2700 seconds, to get the energy in ergs.  This gives us lower limits on the energy radiated during the 45 minute exposures of $>$5.2$\times$10$^{40}$ and $>$6.8$\times$10$^{40}$ ergs, for the two superflares.

The fifth calculation is to correct from the flux recorded in a 45 minute exposure to the total flux of the superflare.  That is, most of the known V2487 Oph superflares have durations longer than 45 minutes, so the plates can only record some fraction of the total flux.  To calculate the total flux, we need to know the light curve shapes for the 1940s superflares, and we do not have this specific knowledge.  Nevertheless, we can place a good upper limit on this fraction, and hence a lower limit on the correction factor to get the total flux.  This limit comes from the known light curve shapes of the $K2$ flares.  For the 8 most luminous superflares (i.e., the best precedents for the 1940s events), I have calculated the fraction of the total flux inside the optimal 45 minute period.  These fractions average to 0.46, with a range of 0.275--0.790.  This means that the observed 45 minute flux should be multiplied by a correction factor equal to 2.2, with a range of 1.3--3.6.  Actually, with the surety that the exposure times are not such to maximize the recorded flux, we can only say that the correction factor is $>$2.2, with a range from $>$1.3 to $>$3.6.  The most conservative constraint is to simply say that the correction factor is $>$1.3.  With this correction, the total radiant energies in the two flares are $>$6.7$\times$10$^{40}$ and $>$8.9$\times$10$^{40}$ ergs, for the two superflares.  These limits can be combined to say that the two Harvard superflares must have each given off energies of $>$10$^{40.8}$ ergs.

These limits are for the most conservative assumptions made at each turn.  That is, the $>$10$^{40.8}$ ergs limit assumes (1) a bremsstrahlung cut-off at 0.1 keV, (2) the light curve shapes were at the fastest observed duration for bright V2487 Oph events, and (3) the specific exposure times for each event maximized the flux.  These choices were made to push the limit to its extreme value.  But this limit is not the best estimate for the superflare energies.  Rather the best estimate of the flare energy comes with (1) a bremsstrahlung cut-off at 1 keV and $f_{\nu} \propto \nu ^{0.10}$ so as to get a bolometric correction of -8.5 mag, (2) the average light curve shape to give a correction factor of 2.2, and (3) non-optimal start/stop times to increase the calculated flux by perhaps a factor of 1.2.  These three factors are the same for the two bursts, resulting in a best estimate that is 39$\times$ larger than the quoted lower limits.  With this, the best estimates for the total radiant energies of the two Harvard superflares are 2.6$\times$10$^{42}$ and 3.5$\times$10$^{42}$ ergs.  

With appropriate accuracy, the two superflares each have energies of 10$^{42.5}$ ergs.  These are enormous amounts of energy for a single event, being 30-billion in units of the Carrington event ($10^{32}$ ergs).  The uncertainty on this is roughly a factor of 2.2$\times$ from the uncertainty of $A_B$, a factor of 39$\times$ from the bolometric correction with its poorly known X-ray cut-off for the bremsstrahlung, and a factor of something like 1.5$\times$ for various possible light curve shapes and timings.  Taken together in quadrature, the log of the energy has an uncertainty of $\pm$1.6.  So the two superflares each have energies of 10$^{42.5\pm1.6}$ ergs.

The superflare energy radiated over, say, one year will depend on the number-distribution.  This distribution is certainly a power law with index close to -2.0.  We have three constraints, from Harvard, the ZTF, and the $K2$, with these all being cumulative measures:  The maximal superflares with energy $>$$10^{42.5}$ ergs occur more frequently than once-a-year.  The $K2$ superflares with red amplitudes of $>$0.1 mag average to once-a-day, with these corresponding to 50$\times$ smaller energy than the 1941 and 1942 events.  For ZTF superflares with amplitudes $>$0.25 mag (see the next Section), with 20$\times$ smaller than the 1941 and 1942 events, the rate is once every 0.44 days.  These three rate estimates have substantial scatter about the expected $E^{-2}$ distribution.  An average gives  $dN/dE$=$4E^{-2}$ superflares per year, for $E$ in units of $10^{41}$ ergs, with an uncertainty of around 3$\times$.  For superflares from $10^{41}$ to $10^{42.5}$ ergs, the total radiated energy in superflares is $10^{42.1}$ ergs in one year.  In the 82 year interval from 1941 to 2023, V2487 Oph has used $10^{44.1}$ ergs.

\section{SUPERFLARES IN ZTF LIGHT CURVES}

The {\it Zwicky Transient Factory} (ZTF) program provides light curves of most stars in the sky (north of $-$28$\degr$ declination) to fainter than 20th magnitude\footnote{\url{https://irsa.ipac.caltech.edu/cgi-bin/Gator/nph-scan?projshort=ZTF}}, typically with over a hundred magnitudes per year since 2018  (Bellm et al. 2019).  For V2487 Oph, with the ZTF data release DR20, we get 277 {\it zg} magnitudes and 532 {\it zr} magnitudes from 2018.2 to 2023.7.  (This excludes magnitudes with non-zero quality flags.)  Of these, 308 magnitudes are taken with fast time series on just four night in the middle of 2018.  Schaefer et al. (2022) has already published the wonderful superflare light curve from one of these nights, with a fast-rise exponential-decay shape with total duration of $>$87 minutes and an amplitude of 0.36 mag in the {\it zr} band.  The usual cadence of these 30 second exposures was to have one-or-two measures per night, with the nights spaced from one day to one week apart, for the observing seasons of half a year each.  For an adopted threshold\footnote{This threshold is for a median {\it zg} of 17.55 and hence a minimum amplitude of 0.25 mag for superflare recognition.  The median {\it zr} is 16.98, so for the same amplitude, the threshold is 16.73 mag.  Any threshold cannot be readily translated to the $K2$ light curves because the background flux has large uncertainties due to the foreground stars inside the $K2$ pixel.} of {\it zg}=17.30 mag, and grouping measures together within 0.05 days, the ZTF data have 22 superflares.  The brightest measure is {\it zg}=16.57, for an amplitude of 0.98 mag above the median brightness level.

The superflare rate can be estimated from the fraction of independent time intervals with brightness measures that have V2487 Oph flaring.  Here, the independent time intervals is taken as 0.05 days, which is roughly the total superflare duration.  For the entire ZTF data set, we have 266 intervals where a superflare would be detected and 22 superflares detected, so the probability is 0.083 that V2487 Oph is flaring at any given time.  So the mean duration brighter than 0.25 mag is 8.3 per cent of the mean flare-to-flare time.  For a mean equivalent duration of 3154 seconds (Schaefer et al. 2022), this makes the average flare-to-flare time of 0.44 days. 

The superflare rate apparently varies from year-to-year.  In 2019, with only 2 observed superflares for 50 independent time intervals, the rate is 4.0$\pm$2.8 per cent.  In 2021, ZTF showed 8 superflares from 66 times, for a rate of 12.1$\pm$4.0 per cent.  This factor of 3$\times$ variation is not highly significant.  Nevertheless, $K2$ already showed variations in the flare rate on shorter timescales, where the first 7 days had 0 superflares, yet the next 62 days had 60 superflares.

From days 7--25 of the $K2$ run, the flare light displayed a periodicity of 1.2216$\pm$0.0031 days, but this periodicity was not displayed by the out-of-flare continuum light over that same interval, and not by the light from days 0--7 and days 25 to the end.  Our large amounts of ground-based time series also showed an apparent periodicity of 1.24$\pm$0.02 days (Schaefer et al. 2022).  So the superflares were occurring with an apparent periodicity of near 1.24 days, with this being a transient effect that comes and goes.  With this history, it is of interest as to whether the ZTF light curve displays a similar periodicity?  A discrete Fourier transform of the ZTF light curve reveals no significant periodicity anywhere near 1.24 day superflare period, nor at the 0.753 day orbital period, nor at any other period.\footnote{A small number of flares occurring randomly within a set time interval will always show a loose clustering in phase for some {\it ad hoc} set of periods (Schaefer \& Desai 1988).  In the ZTF light curve, such weak clusters vary greatly in period with the exact choice of time interval, and are not significant.}  When each observing season is examined separately, no significant periodicity is seen.  So the ZTF data do not confirm or display the transient 1.24 day periodicity as seen for some of the $K2$ flares and as seen weakly from the many ground-based time series.

\section{RESULTS, ANALYSIS, AND LOOKING FORWARD}

The primary results of this paper are to extend the extreme properties of V2487 Oph.  Two superflares have been measured with total radiative energies that are confidently $>$6.7$\times$10$^{40}$ and $>$8.9$\times$10$^{40}$ ergs, while the best estimates are 2.6$\times$10$^{42}$ and 3.5$\times$10$^{42}$ ergs.  Both events have energies of $10^{42.5\pm1.6}$ ergs, or $\sim$30-billion Carringtons.  Further, the duration of the superflare `stage' has been extended to 1941--2023, so this is not any type of a short `high-state'.   For the measured number-distribution, the best estimate for the total energy radiated by superflares over this 82 year interval is $\sim$$10^{44.1}$ ergs.

What is the ultimate source for this enormous energy?  Well, the proximate source for the superflare energy is from the electron kinetic energy as they smash into their own footprints to radiate by the bremsstrahlung process, with this electron energy coming from the reconnection of the magnetic field.  So where does the energy come from that is powering the magnetic field?  The obvious source is that the kinetic energy of the binary orbit is somehow getting extracted and turned into magnetic field energy.  Still, it is worthwhile to consider other sources for the energy reservoir that powers the magnetic fields:  (1) The total kinetic energy of the binary, with a 1.35 $M_{\odot}$ white dwarf, a 0.96 $M_{\odot}$ companion star, and a 0.753 day orbit, is $10^{47.7}$ ergs.  This is easily large enough to power the superflares.  (2) The companion's nuclear burning will produce energy at a rate of near 3$\times$10$^{33}$ erg per second, for $10^{41.0}$ ergs in a year.  This energy source is too small to power the superflares.  (3) The magnetic field of the companion star is unknown, but isolated main-sequence stars with radii around 0.96 $R_{\odot}$ have surface magnetic fields only up to values like 1000 Gauss.  For a dipole field with 1000 Gauss at the stellar surface, the magnetic energy external to the star is $10^{37.7}$ erg.  This is greatly too small to power the superflares.  (4) The accretion energy for an accretion rate of $10^{-7}$ $M_{\odot}$ yr$^{-1}$ is $10^{44.1}$ ergs in one year.  So the superflares can be powered by tapping only one percent of the accretion energy.  (5) The spin period of the white dwarf will be greatly slower than the Keplerian value due to the angular momentum loss during each nova event, with typical values of 100--1000 seconds for both nonmagnetic dwarf novae and intermediate polars (Livio \& Pringle 1998).  For a spin period of 100 seconds on a 1.35 $M_{\odot}$ white dwarf, the kinetic energy of rotation is $10^{47.1}$ ergs, which is enough to power superflares for $\sim$$10^5$ years.  (6) The surface magnetic field of the white dwarf is $<$$10^6$ Gauss, so the external magnetic energy is $10^{36.6}$ ergs, which is too small to power the superflares.  In all, there are three energy reservoirs that are large enough to power the superflares, with these being the stars' kinetic energy in their orbits, the gravitational energy from the high accretion rate, and the white dwarf's kinetic energy of rotation.

{\it If} the superflare energy is ultimately extracted from the kinetic energy of the orbit, then each superflare would shorten the orbital period, decrease the size of the companion's Roche lobe, and drive material to overflow off the star.  I calculate that a single 10$^{42.5}$ ergs superflare would make the companion's Roche lobe radius shrink by 6.6 km.  (This is for adopted stellar masses of 1.35 M$_{\odot}$ for the white dwarf and 0.96 M$_{\odot}$ for the companion.)  And that is just for one superflare.  I do not know the effective density of the companion's stellar atmosphere near the inner Lagrangian point, so I cannot calculate the induced accretion rate.  For the observed number distribution (see Section 5), the Roche lobe will shrink nearly 2.9 km per year.  In the time from 1941 to 2023, the Roche lobe would shrink by $\sim$240 km.  This calculation has substantial uncertainty, both from the number-distribution and the bremsstrahlung X-ray cutoff.  This change in the Roche lobe radius can be compared to the atmospheric scale height of 150 km (for a main sequence star with a size of 0.96 R$_{\odot}$).  (Indeed, this shrinkage is so large that there must be some substantial inefficiency so as to avoid the accretion rate and $B_q$ changing greatly since 1941.)  With this, the superflare energy depletion of the orbit is seen to dominate the driving of the accretion, and being adequate to power the high rate of V2487 Oph.  However, before any evolution calculations or generalizations or conclusions are made, we must have a clear understanding of how the energy for the magnetic field lines is drawn from the binary orbit.

With V2487 Oph having long-running near-daily superflares, ongoing even up to recent months, this opens up the possibility of monitoring programs being able to reliably capture the superflare.  Rodr\'{i}guez-Gil et al. (2023) used this tactic to get the all-time first optical spectrum of any superflare.  So now, we know that we can get many more superflare spectra, just by sitting on V2487 Oph for up to 24 hours of telescope time.  It will be wonderful to get a good time series of spectra showing the changes in line profiles and lines and velocities throughout a superflare.  It will be great to get many spectra, for example to plot out the ejection velocities as a function of orbital phase.  For the issue raised in this paper regarding the bolometric correction, a stare with an X-ray satellite for a day or two should likely catch the high energy spectrum, hence defining the bolometric correction, the emission mechanism, and the superflare properties.  A good time for this long X-ray stare would be during April and May 2025, when {\it TESS} is scheduled to monitor V2487 Oph full-time during Sectors 91 and 92.

V2487 Oph now represents the extreme case for two separate phenomena, novae and superflare stars.  As a nova, V2487 Oph is the case of extremely fast recurrence times, being one of the fastest known in our Galaxy.  As a superflare star, V2487 Oph has by-far the fastest recurrence timescale (roughly once a day) and the highest energy ($\sim$10$^{42.5}$ ergs).  Such a pair of extremes in the same system is unlikely unless the properties are causally connected.  

So, why does V2487 Oph represent extreme cases for both novae and superflare stars?  To have an extreme nova recurrence timescale, the CV must have its white dwarf mass close to the Chandrasekhar limit and its accretion rate up near the steady hydrogen burning limit.  To have the fastest and highest-energy superflares, I think that the system need only have an accretion disk serving as footprints for magnetic loops connecting the star to the disk.  But these requirements for extreme novae and superflares are not causally connected, so I do not understand their simultaneous operation in any one binary.  

A related similar question is to ask why U Sco and CI Aql (two very-well-observed RN sisters of V2487 Oph) do not have superflares, much less extreme superflares?  All three RN sisters have essentially the same orbital period, near-Chandrasekhar-mass white dwarfs, companions of roughly the same radius, mass, and temperature as our Sun, and accretion into a disk with rates near the maximum.  So the deeper reason why V2487 Oph has extreme superflares must be not readily visible.  For this, the obvious idea is that V2487 Oph has very large magnetic fields for some unknown reason, while its sister-RNe have no significant fields.  Still, this is all to say that we have no useable idea as to {\it how} or {\it why} V2487 Oph has such extreme superflares.  These higher-level questions of {\it how} and {\it why} are now the forefront for both V2487 Oph research and superflare research.

\begin{acknowledgments}
This work has made use of data provided by Digital Access to a Sky Century @ Harvard (DASCH), which has been partially supported by NSF grants AST-0407380, AST-0909073, and AST-1313370.  I am thankful for the ZTF providing excellent light curves of most stars in the sky north of roughly $-$28$\degr$ declination.
\end{acknowledgments}

%

\vspace{5mm}
\facilities{DASCH, ZTF, APASS}







{}


\end{document}